\documentclass[12pt]{article}
\usepackage{latexsym, amsbsy, amssymb,multirow, epsfig, amsmath}
\usepackage{hyperref}

\usepackage{natbib,graphicx,setspace,lscape,longtable}
\usepackage{natbib,epsfig,graphicx}

\usepackage{amsmath,amsthm,amssymb,color}

\RequirePackage[mathlines, displaymath]{lineno}
\bibpunct{(}{)}{;}{a}{,}{,}

\newcommand{\csection}[1]
    {\begin{center}
        \stepcounter{section}
        {\bf\large\arabic{section}. #1}
    \end{center}
    \vspace{-0.15 cm}
}

\newcommand{\scsection}[1]
    {\begin{center}
        {\bf\large #1}
    \end{center}
    \vspace{-0.15 cm}
}

\def\beqr{\begin{eqnarray}}
\def\eeqr{\end{eqnarray}}
\def\beqrs{\begin{eqnarray*}}
\def\eeqrs{\end{eqnarray*}}
\def\bep{\begin{prop}}
\def\eep{\end{prop}}

\numberwithin{equation}{section}

\newtheorem{theo}{\bf Theorem}[section]

\newtheorem{prop}{\bf Proposition}[section]

\newtheorem{lemma}{Lemma}

\newtheorem{remark}{Remark}

\usepackage{latexsym}
\usepackage{epsfig}
\usepackage{bm}

\usepackage{threeparttable}
\usepackage{graphicx}
\usepackage[small]{caption2}
\usepackage{threeparttable}
\usepackage{dcolumn}
\usepackage{multirow}
\usepackage{booktabs}
\usepackage{footnote}
\usepackage[linesnumbered,boxed]{algorithm2e}
\usepackage[final]{pdfpages}

\DeclareSymbolFont{largesymbol}{OMX}{yhex}{m}{n}
\DeclareMathAccent{\Widehat}{\mathord}{largesymbol}{"62}

\begin{document}

\begin{center}
{\bf\large Detecting multiple change points: a PULSE criterion}
\\ \vskip0.6cm
Wenbiao Zhao$^1$, Xuehu Zhu$^2$ and Lixing Zhu$^{3, 4}$\footnote{The research was supported by a grant from the University Grants Council of Hong Kong. } 
\\
\textit{ $^1$ Renmin University of China\\
$^2$ Xi'an Jiaotong University\\
$^3$ Beijing Normal University\\
$^4$ Hong Kong Baptist University }

\today
\end{center}

\begin{singlespace}
\begin{abstract}
	The research described herewith investigates detecting change points of means and of variances in a sequence of observations. The number of change points can be divergent at certain rate as the sample size goes to infinity. We define a MOSUM-based objective function  for this purpose. Unlike all existing MOSUM-based methods, the novel objective function exhibits an useful ``PULSE" pattern near change points in the sense: at the population level, the value at any  change point plus 2 times of the segment length of the moving average attains a local minimum tending to zero following by  a local maximum going to  infinity.  This feature provides an efficient way to simultaneously  identify all change points at the sample level.  In theory,  the number of change points can be consistently estimated and  the locations can also be consistently estimated in a certain sense. Further, because of  its visualization nature, in practice, the locations can be relatively more easily identified  by plots than existing methods in the literature. The method can also handle the case in which the signals of some change points are very weak in the sense that those changes go to zero.  Further, the computational cost is very inexpensive.  The numerical studies we conduct validate its good performance.
\end{abstract}


\end{singlespace}

\newpage
\csection{Introduction}
Change points detection  has attracted significant attention in a variety of research fields in decades (see, e.g. \citet{ref19}). There are a number of methods available in the literature to detect sudden structure changes at certain points, that is, change points. For example \citet{ref2} detected changes of mean in time series data for financial modeling, and \citet{ref10} identified genes associated with some diseases by applying a method of change points detection  for means.

Methods designed for detecting change points in mean or variance with unknown number have been proposed in the literature. Almost all existing approaches for multiple change points could be roughly divided into two prevalent categories: model selection-based and hypothesis testing-based. For model selection-based approaches, as an example, \citet{ref16} firstly suggested a BIC type criterion for this purpose. More recently, regularization-based optimization approaches   have been proposed. \citet{ref17} proposed a penalized least squares-based approach for mean changes. A weighted least squares function-based method was suggested by \citet{ref7}. \citet{ref18} proposed a LASSO-based approach. { These approaches have been used in the cases with the fixed number of change points to obtain consistent estimators.}
For hypothesis testing-based methods, when detecting changes in means or variances, to facilitate the testing procedure by bisection algorithm  a cumulative sum-based approach (CUSUM) that was firstly proposed by \citet{ref19}  has become a cornerstone in the later developments in such  methodologies.
\citet{ref20} designed some tests for multiple changes through binary segmentation methods.  To alleviate this difficulty caused by  short spacings between change points or  small jump magnitudes, \citet{ref4} introduced an additional randomization step in the algorithm.
Moving sum (MOSUM) or scan statistics are also popularly used to  construct tests such as \citet{ref21} and \citet{ref22}. {In addition, \citet{mocum} applied this methodology to handle multivariate time series data.} Recently, \citet{ref2} and \citet{ref1} discussed  the limiting distributions of the maxima of MOSUM. The corresponding  tests can well control their sizes, under the null hypothesis, close to the significance level.

Both classes of methodologies are usually efficient in estimation. As for their limitations,  regularization-based estimations involving optimization algorithms have the problem of computational complexity, there are no results about the cases with divergent number of change points as the sample size goes to infinity. While  hypothesis testing-based estimations have to  benefit from bisection procedures to define test statistics for sequential testing,  the resulting estimators are not consistent. The cases with divergent number of change points have not yet been discussed either. Using such procedures is partly because these methods do not have an implementable objective function to define a criterion to simultaneously detect all change points.

In this paper, we propose a novel  approach for this purpose. Motivated by the idea of MOSUM, we define an objective function via a sequence of ridge ratios of  moving averages.  Note that to well identify change points, the key is how to make the values at the true change points(or nearby in certain sense) stand out. 
Unlike existing MOSUM-based methods and loss-function-based methods as well, the most distinguishing feature of the new criterion we will propose is that the defined objective function is  discontinuous with an useful ``PULSE" pattern near all change points: at the population level any change point plus 2 times of the segment length of moving average attains a local minimum tending to zero following by a local maximum going to infinity. Thus, this feature can very much make change points stand out and thus provide an efficient way to identify them. We will give a toy example to show this pattern in Section~2 when we describe the criterion construction. 
 It is worthwhile to mention that because of the visualization nature, the plot of objective function can  make all change points visualized and thus in practice it is very easy to implement with the help from plot. It is also computationally inexpensive without involving any optimality algorithm. We call this method a PULSE criterion. To show its usefulness, We will check how sensitive the criterion is to ``weak changes" in the sense that some changes in  the sequence of local means converge to, at a certain rate, a sequence without mean changes. {As a generic methodology, it could be  extended to handle other change points detection problems such as distributional changes (e.g. \citet{pollak}), changes in regression models (e.g. \citet{refred72}) and change points of functional data (e.g. \citet{functional}). The research is ongoing.} We also understand that it has the limitation to handle the problems with  short spacings between two change points. This is because, to guarantee the estimation consistency, the segment length of the moving averages needs to be sufficiently large, which would contain more than one change point. We will have a brief discussion in Section~6.



The paper is organized as follows. In Section~2, we introduce the criterion construction, and investigate the estimation consistency. Section~3 contains the investigation on weak signals case where the magnitudes of the changes converge to zero at a certain rate. As an extension, Section~4  includes the detection of changes  over variances. Some numerical studies are put in Section~5. Section~6 contains an illustrative application to the detection of mean changes. Section~7 includes some discussions for the merits and limitations of the method and some further research topics. All the technical proofs are presented in  Appendix.

\csection{Methodological Development}
\subsection{Notations}
 Let $X_1,..., X_n$ be independent one-dimensional random variables as
\begin{center}
$X_i = \mu_i + \varepsilon_i,     1\leq i \leq n $,
\end{center}
where $\mu_i=E(X_i)$ are the means.
Assume that the sequence of all means follows a piecewise constant structure with $K+1$ segments. In other words, there are $K$ change points
$1 < z_1 < z_2 <...< z_K< n$
such that
$\mu_{z_{k-1}+j} =\mu^{(k)}$, for $k=1, \cdots, K+1$ and $0\leq j\leq z_{k}-z_{k-1}-1$
where  $z_0=0$ and $z_{K+1}=n$. For $k=1,...,K$, write
$\beta_k=|\mu^{(k+1)}-\mu^{(k)}|$ for the (non-zero) difference in means between consecutive segments. The number $K$  can go to infinity as the sample size goes to infinity.

The goal of this research is to suggest a detection method for the locations of change points $\{z_1,...,z_{K}\}$ in the data stream and  the number $K$. To well estimate these consistently in a certain sense, we give some assumptions on the magnitudes of the mean changes $\{\beta_{k}: 1\leq k \leq K\}$ and the lengths  $z_{k+1}-z_k$ of segments. For notational convenience, write the minimum length of segments satisfies  as $\alpha^*$:
\begin{eqnarray}\label{min_length}
\alpha^*:=\min_{0\le k\le K}  \{ z_{k+1}-z_{k} \}
\end{eqnarray}
and the minimum magnitudes of mean changes  as $\nu$:
\begin{eqnarray}\label{difference}
\nu:=\min_{1\le k\le K} \beta_{k}.
\end{eqnarray}
 Write  $1 < \hat z_1\le \hat z_2\le...\le \hat z_{K}< n-1$ as the estimated locations.

\subsection{Criterion Construction}

To achieve the goal discussed above,  an objective function is constructed by the following steps. Consider the mean changes detection problem.

{\it Difference of Moving Averages:} To character the mean information,  let $S(i)$ be the moving sum with length of $\alpha_n$ for every location $i$ as:
\begin{equation}
S(i)=\sum_{j=i}^{i+\alpha_n-1}\mu_j
\end{equation}
where $\alpha_n$ is called the moving window.
As the difference between two successive moving sums at the population level can show the mean change at its location $z_k$, we define $D(i)$ as: for $1\leq i \leq n-2\alpha_n,$ if $2\alpha_n< \alpha^*$,
\begin{eqnarray}
 D(i)&:=&{1\over \alpha_n}(S(i)-S(i+\alpha_n))\nonumber\\
&=&{1\over \alpha_n}(\sum_{j=i}^{i+\alpha_n-1}\mu_j-\sum_{j=i+\alpha_n}^{i+2\alpha_n-1}\mu_j)
\end{eqnarray}
For any fixed $k$, we have:
\begin{eqnarray}
D(i)=\left\{
\begin{array}{ll}
 {i-(z_k-\alpha_n)\over \alpha_n}(\mu_{k+1}-\mu_k), & z_k- \alpha_n \leq i < z_k, \\
{z_k+\alpha_n-i\over \alpha_n}(\mu_{k+1}-\mu_k), & z_k\leq i \leq z_{k}+\alpha_n\\
0, & z_{k-1}+\alpha_n\leq i\leq z_k-\alpha_n.
\end{array}
\right.
\end{eqnarray}
This is because, when $z_{k-1}+\alpha_n\leq i\leq z_k-\alpha_n$, $S(i)=S(i+\alpha_n)$.
 $D(i)$ attains  a local maximum/minimum at $i=z_k$ for any $k$ with $ 1\leq k\leq K$ within the interval of length of $2\alpha_n$. Figure~1 presents the plot for  visualizing the pattern. This is not a new idea while just the idea of MOSUM.  However, as the local maximum/minimum values cannot be  quantified  to form an useful objective function, and more seriously,  at the sample level, we can expect too many local maxima/minima due to the randomness oscillation, only identifying them makes no sense for the change points detection.  Thus, we consider further construction to define an implementable objective function. Our idea is to construct a sequence of ridge ratios as an objective function that is of a pulse pattern  
 within the space between a local maximum of $|D(i)|$ tending to infinity and a local minimum going to $0$ is $\alpha_n$ at the population level.
 However,  it is easy to understand that as $D(i)$ and then the objective function at the sample level  is relatively oscillating,  it may cause difficulty to accurately determine the number of change points and locations. To make an objective function more smoothly at the sample level, we consider smoother averages than $D(i)$'s by doubly averaging below.

 {\it Doubly Averaging:} The second round of averaging is to repeatedly use  datum points in every average. To further make the values near the true change points stood out and more importantly, to establish a criterion for implementation, we use an idea of ridge ratio to construct an objective function. Denote $\tilde D(i)$ by the averages of $D(i)$ within the segments of length $\alpha_n$:
\begin{equation}
\tilde D(i)={1\over \alpha_n}\sum_{j=i}^{i+\alpha_n-1}D(j).
\end{equation}
As the result, we have the following properties:
\begin{eqnarray*}
	|\tilde D(i)|=\left\{
	\begin{array}{lll}
		0, &z_{k-1}+\alpha_n \leq i\leq z_k-2\alpha_n; \\
		
		{(i-z_k+2\alpha_n+1)\cdot(i-z_k+2\alpha_n) \over \alpha_n^2}\beta_k, &z_k-2\alpha_n<i\leq z_k-\alpha_n; \\
		
		{-i^2-\alpha_ni+2iz_k-i+z_k-z_k^2+\alpha_nz_k+{1\over 2}(\alpha_n^2-\alpha_n)\over \alpha_n^2}\beta_k,
		& z_k-\alpha_n<i < z_k-{\alpha_n\over 2}-{\sqrt \alpha_n}; \\
		
		({3\over 4}-{\sqrt\alpha_n-1\over \alpha_n})\beta_k, &i=z_k-{\alpha_n\over 2}-{\sqrt \alpha_n}; \\
		
		{-i^2-\alpha_ni+2iz_k-i+z_k-z_k^2+\alpha_nz_k+{1\over 2}(\alpha_n^2-\alpha_n)\over \alpha_n^2}\beta_k,
		& z_k-{\alpha_n\over 2}-{\sqrt \alpha_n}<i < z_k-{\alpha_n\over 2}; \\
		
		{3\over 4}\beta_k, &i=z_k-{1\over 2}\alpha_n; \\
		
		{i^2+\alpha_ni-2iz_k+i-z_k+z_k^2-\alpha_nz_k-{1\over 2}(\alpha_n^2-\alpha_n)\over \alpha_n^2}\beta_k, &z_k-{\alpha_n\over 2} < i < z_k-{\alpha_n\over 2}+{\sqrt \alpha_n}; \\
		
		({3\over 4}-{\sqrt\alpha_n-1\over \alpha_n})\beta_k, &i=z_k-{\alpha_n\over 2}+{\sqrt \alpha_n}; \\
		
		{i^2+\alpha_ni-2iz_k+i-z_k+z_k^2-\alpha_nz_k-{1\over 2}(\alpha_n^2-\alpha_n)\over \alpha_n^2}\beta_k, &z_k-{\alpha_n\over 2} +{\sqrt \alpha_n} < i \leq z_k; \\
		
		{(-i+z_k+\alpha_n+2)(-i+1+\alpha_n+z_k) \over \alpha_n^2}\beta_k, &z_k< i \leq z_k+\alpha_n; \\
		
		0, &z_k+\alpha_n<i\leq z_{k+1}-2\alpha_n.
	\end{array}
	\right.
\end{eqnarray*}
where $\beta_k=|\mu_{k+1}-\mu_k|$.
Thus,
\begin{eqnarray*}
      \tilde D(i)\left\{
                    \begin{array}{ll}
                    > 0, & z_k- 3\alpha_n \leq i \leq z_k, \\
                    =0, & otherwise.
                    \end{array}
                  \right.
      \end{eqnarray*}
Clearly, $\tilde D(i)$ attains local maxima at $z_k-{1\over 2}\alpha_n$ for each $k$ with $1\leq k\leq K$.  The local maximizers of  $\tilde D(i)$ plus ${1\over 2}\alpha_n$ are the locations of change points. Similarly as $D(i)$, the sequence $\tilde D(i)$ cannot be directly used to be an objective function either. Now we construct a sequence of ridge ratios as  an objective function that is of a ``pulse" pattern  such that change points can be well identified.

{\it Objective Function}. \,
Consider the ratios between $\tilde D(i)$ and $\tilde D(i+{3\over 2}\alpha_n)$. 
Define  the ridge ratios $T(i)$ at the population level as
\begin{equation}
T(i)={{|\tilde D_(i)|+c_n} \over {|\tilde D(i+{3\over2}\alpha_n)|}+c_n},
\end{equation}
where $c_n \to 0$ is a small value, to be selected later, to avoid the unstable terms $0/0$. In addition, for $i\in(z_{k-1}+\alpha_n,z_k-2\alpha_n)$, $|\tilde D(i)|=0$ and $|\tilde D(i+{3\over 2}\alpha_n)|$  monotonically increases. For $i \in (z_k-2\alpha_n,z_k-{1\over 2}\alpha_n)$, $|\tilde D(i))|$ monotonically increases, and $|\tilde D(i+{3\over 2}\alpha_n)|$ monotonically decreases. For $i\in (z_k-{\alpha_n\over 2}, z_k+\alpha_n)$, $|\tilde D(i+{3\over 2}\alpha_n)|=0$ and $|\tilde D(i)|$ monotonically decreases. Then $c_n$ could also play a role of making $T(i)$ monotonic, to avoid the scenario where there are too many points  tending to $0$.
In summary, the following property could be easily justified: letting $\searrow$ and $\nearrow$ mean decreasing and increasing with respect to the index $i$; $\to 0$ and $ \to \infty$ mean going to  zero and infinity as the sample size $n \to \infty$,
\begin{eqnarray*}
      T(i)=\left\{
                    \begin{array}{ll}
                    1, &z_{k-1}+\alpha_n\leq i \leq z_k-{7\over 2}\alpha_n, \\
                     {c_n\over {|\tilde D(i+{3\over2}\alpha_n)|+c_n}} \searrow , & z_k-{7\over 2}\alpha_n<i < z_k-2\alpha_n, \\
                      {c_n\over {|\tilde D(i+{3\over2}\alpha_n)|+c_n}} \to 0, & i = z_k-2\alpha_n, \\
                     {|\tilde D(i)|+c_n\over |\tilde D(i+{3\over2}\alpha_n)|+c_n} \nearrow , & z_k-2\alpha_n<i< z_k-{\alpha_n\over 2}, \\
                     {{|\tilde D(i)|+c_n}\over c_n} \to \infty , &i=z_k-{\alpha_n\over 2},\\
                       {|\tilde D(i)|+c_n\over c_n} \searrow , & z_k-{\alpha_n\over 2}<i< z_k+\alpha_n, \\
                    1, &z_k+\alpha_n \leq i < z_{k+1}-{7\over 2}\alpha_n.
                    \end{array}
                  \right.
      \end{eqnarray*}
Any true change point is just the index of a local minimum plus $2\alpha_n$. Based on this objective function,  using the local minimizers to identify change points is  convenient to implement.
The  toy example in Figure~1 shows the curve patterns of $D(i)$, $|\tilde D(i+\alpha_n/2)|$ and $T(i+ 2\alpha_n)$ such that we have a better idea to understand why the pulse pattern of the objective function $T(i)$, rather than that of $D(i)$ or of $|\tilde D(i)|$, can be used to construct an useful criterion. Based on these properties, we can  define its empirical version.
\begin{figure}
	\centering
	\includegraphics[width=5.5in]{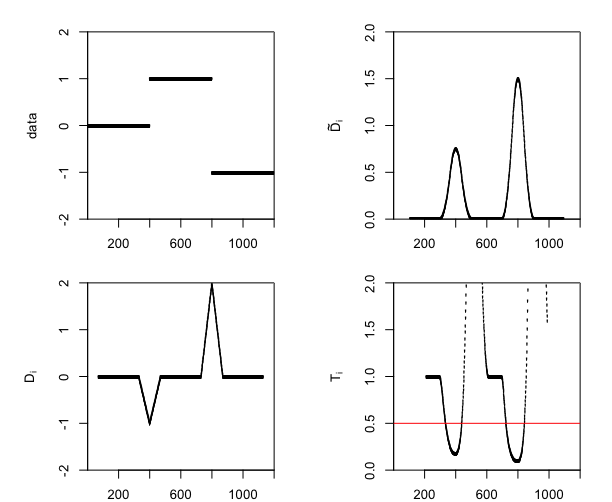}
	\caption{The plots at the population level}
	\label{picpopu}
\end{figure}


\textit{Sample Version.}
To define the objective function at the sample level, we can use the sample averages to estimate  $D(i)$ and $\tilde D(i)$. Denote $\hat S(i)=\sum_{j=i}^{i+\alpha_n-1}X_j$ as the estimator of  $S(i)$, $D_n(i)={1\over \alpha_n}(\hat S(i)-\hat S(i+\alpha_n))$ and $\tilde D_n(i)={1\over \alpha_n}\sum_{j=i}^{i+\alpha_n-1}D_n(j)$. The sample version of the objective function is then defined as: for $i=1,.., n-{7\over 2}\alpha_n$,
\begin{equation}
T_n(i)={{|\tilde D_n(i)|+c_n} \over {|\tilde D_n(i+{3\over2}\alpha_n)|}+c_n},
\end{equation}
and the ridge value $c_n$ tends to $0$ at a certain rate specified later.  We can see that $E\tilde D_n (i)= \tilde D(i)$. 

{\it Criterion:} It is understandable that  at the sample level, the objective function should be highly oscillating and there would be too many local minima. Thus, a natural idea is to restrict the search such that  within each chosen interval, there is only one minimum of $T_n(i)$. We do this through a  threshold $\tau$ with $0<\tau<1$. That is:
\begin{center}
$\{ i, 1 \leq i \leq n-{7\over 2}\alpha_n: T_n(i)< \tau \}$.
\end{center}
To make the search easily, we simply recommend $\tau=0.5$ as a compromised value as for large $\tau$ close to $1$, there would exist more local minima and for small $\tau$ close to $0$, there would exist less even no local minima.
From the properties of $T_n(i)$ that can also be seen from the plot of Figure~1 heuristically, we can see that all these indices can be separated into several disjoint subsets each containing only one change point asymptotically. Therefore, we can separately within the disjoint subsets  search for local minima.  From the definition of $T(i)$ at the population level, we can see that the gap between two local minimizers must be larger than $2\alpha_n$. Due to the consistency of the involved estimators, we can see that there are $\hat K$ pairs $(m_k, M_k)$  where $m_k$ and $M_k$ are determined by $T_n(i)< 0.5$ and $m_k$ satisfies that $T_n(m_k-1)\ge 0.5$ and $T_n(m_k)< 0.5$, and $T_n(M_k)< 0.5$ and $T_n(M_k+1)\ge  0.5$. Write $\hat z_k-2\alpha_n$ as the minimizer in each interval.  

\begin{theo}\label{theo1}
	Assume that $X_i-EX_i$ are independent identically distributed random variables and ${\alpha^*\over {n^{1/2}(\log n)^2 }}\rightarrow \infty$, where $\alpha^*$ is the minimum number of samples between any two change points. The tuning parameter $c_n$ and the segment length $\alpha_n$ satisfies that ${c_n\over  {\sqrt{\log n\over \alpha_n}}}\rightarrow \infty$, ${ n^{1/4} \log n\over {\sqrt\alpha_n}}\rightarrow 0$, and ${\alpha^*\over \alpha_n}\rightarrow \infty$.
	
	(1) When $K$ is known, then the estimators $ \{\hat z_1,..., \hat z_K\} $ have: $\lim_{n \rightarrow \infty} \Pr\{ \max_{1\le k\le K}{|{\hat z_k-z_k\over \alpha_n}|}<\epsilon\}\rightarrow1$, for every $\epsilon>0$.
	
	(2) When  $K$  is fixed but unknown, then $\hat K= K$ with a probability going to one and the estimators $ \{\hat z_1,..., \hat z_{\hat K}\} $ have: $\lim_{n \rightarrow \infty}\Pr\{ \max_{1\le k\le K}{|\hat z_k-z_k|\over \alpha_n}<\epsilon\}\rightarrow1$, for every $\epsilon>0$.
	
	(3) When $K$ is divergent at the rate satisfying {\color{red} ${n\over \alpha^* K}\rightarrow \infty$} and unknown, the results are the same as those in (2). 	
\end{theo}

\begin{remark}
These conditions are based on the following observations. First, we will prove that 	$\tilde D_n(i)-\tilde D(i)$ converges to $0$ at a rate of order $\sqrt{\log n\over \alpha_n}$. Then the ridge $c_n$ should be a dominating term in every  $T_n(i)$,  which converges to $0$ at a rate slower than that of $\tilde D_n(i)-\tilde D(i)$. Such a ridge can help $T_n(i)$ keep the property of $T(i)$ asymptotically. We will show this in the proofs of the theorems in Appendix.
	
	In the estimation,  the ridge $c_n$ needs to be selected.  We will recommend a choice for practical use in the numerical studies. To guarantee the estimation consistency,	 the bandwidth $\alpha_n$ should not be too small such that the averages can be close to the corresponding means. Thus, for the paradigms with very short spacing, our method, similarly as any MOSUM-based methods, may not perform well. This deserves a further study.
\end{remark}

\begin{remark}
	In the third part of this theorem, we allow the number of change points divergent. To the best of our knowledge, there are not many proposals in the literature to discuss this problem. \citet{zounum} developed a cross-validation estimation scheme, the estimation consistencies of the number and  locations of change points were not provided. In contrast, the proposed criterion is of visualization nature and without resampling approximation, the determination is rather convenient and computationally inexpensive.
\end{remark}

\csection{Weak signals case}

In this section, we extend the criterion to handle weak signal scenarios. The term ``weak signals"  in this section means that the magnitudes of some changes  converge to $0$ at a certain rate as the sample size goes to infinity. We also call such models as local models. Consider the sequence of models as, for $1\le k\le K$:
\begin{equation}
X_i=\mu+\beta_{z_k} \mathbb I\{ i\ge z_k \}+\epsilon,  \label{3.1}
\end{equation}
where $z_k$  are the locations of  change points. $\beta_{z_k}$ are the change magnitudes, which  would converge to $0$ as $n$ goes to infinity. Denote $\beta_z=\min_{1\leq k\leq K}\beta_k$. We have the following results.

\begin{theo}\label{theo2}
	Assume that $X_i-EX_i$ are independent identically distributed random variables and ${\alpha^*\over {n^{1/2}(\log n)^2 }}\rightarrow \infty$. The tuning
parameter $c_n$ and $\alpha_n$ satisfies that ${c_n\over  {\sqrt{\log n\over \alpha_n}}}\rightarrow \infty$, ${ n^{1/4} \log n\over {\sqrt\alpha_n}}\rightarrow 0$, and ${\alpha^*\over \alpha_n}\rightarrow \infty$.
	For the sequence of local models in (\ref{3.1}), under the conditions in Theorem~\ref{theo1},  if $ {\log \alpha_n}^{1/5}\beta_z \to \infty$, we have  $\lim_{n \rightarrow \infty}\Pr\{ \hat K=K\}=1$ and   $\lim_{n \rightarrow \infty}\Pr\{ \max_{1\le k\le K}{|\hat z_k-z_k|\over \alpha_n}<\epsilon\}=1$, for every $\epsilon>0$.
\end{theo}


\csection{Extension}
In this section, we extend the criterion to handle  detecting change points in variances. Consider second moments of $X_i$'s that are generated from the following model:
\begin{equation}
X_i = \mu + \varepsilon_i,     1\leq i \leq n ,
\end{equation}
where $\mu$ is an unknown mean and $E(\varepsilon)=0$, $Var(\varepsilon)=\sigma^2_{(i)}$. Similarly, we assume that the variances follow a piecewise constant structure with $K+1$ segments. In other words, there are $K$ change points
$1 < z_1< z_2<...< z_K < n-1$
such that, for any $k$ with  $0\leq k\leq K$,
\begin{equation}
Var(\varepsilon_{z_k+1})=...=Var(\varepsilon_{z_{k+1}})=\sigma^2_{k}.
\end{equation}
As before, define $z_0=0$ and $z_{K+1}=n$. At population level, we can similar define $D(i)$ and $\tilde D(i)$ as follows.
	\begin{equation*}
	D(i)=\sigma_{(i)}-\sigma_{(i+\alpha_n)}
	\end{equation*}
	and
	\begin{equation*}
	\tilde D(i)={1\over \alpha_n}\sum_{j=i}^{i+\alpha_n-1}D(j).
	\end{equation*}

Similarly, at the sample level, as the mean is unknown but all the same, we can estimate $\mu$ by sample mean.

Let\begin{equation}
\hat \sigma^2_{(i)}={1\over \alpha_n}\sum_{t=i}^{t=i+\alpha_n-1}(X_t-{1\over n}\sum_{j=1}^{j=n}X_j)^2.
\end{equation}
and define similarly $D_n(i)$ and $\tilde D_n(i)$ as the difference of moving averages  and the average of $D_n(j)$'s:
\begin{equation}
D_n(i)=\sqrt{\hat \sigma^2_{(i)}}-\sqrt{\hat \sigma^2_{(i+\alpha_n)}},
\end{equation}
\begin{equation}
\tilde D_n(i)={1\over \delta_n}\sum_{j=i}^{i+\delta_n}D_n(j).
\end{equation}
Finally, we take ratios of $\tilde D(i)$ to acquire the statistics,$T(i)$ we will use to estimate change points. That is:
\begin{equation}
T_n(i)={{|\tilde D_n(i)|}+c_n \over {|\tilde D_n(i+{3\over2}\alpha_n)|}+c_n}.
\end{equation}

\begin{theo}\label{theo3}
		Assume that $X_i-\mu$ are independent  distributed random variables and $\alpha^*$, $c_n$ and $\alpha_n$ satisfy the same conditions as in Theorem~\ref{theo1}.
	
    (1) When $K$ is given, then the estimators $ \{\hat z_1,..., \hat z_K\} $ have: $\lim_{n \rightarrow \infty} \Pr\{ \max_{1\le k\le K}{|{\hat z_k-z_k\over \alpha_n}|}<\epsilon\}\rightarrow1$, for every $\epsilon>0$.

    (2) When $K$ is unknown but fixed, then $\hat K=K$ with a probability going to one and the estimators $ \{\hat z_1,..., \hat z_{\hat K}\} $ have: $\lim_{n \rightarrow \infty}\Pr\{ \max_{1\le k\le K} {|\hat z_k-z_k|\over \alpha_n}<\epsilon\}\rightarrow1$, for every $\epsilon>0$.

    (3) When  $K$ is divergent at the rate satisfying {\color{red} ${n\over \alpha^* K}\rightarrow \infty$} and unknown, the results are identical to those in (2).
\end{theo}

\begin{remark}
	This theorem told us that our method could be extended to scenarios for detecting change point in variance. The only difference between mena change scenario and variance scenario is that the former is based on first monment while the later is based on second monment. The proof of this result is very similar to that for mean. Some conditions on forth monment of the sample will also be considered.
\end{remark}

\csection{Simulation}
To evaluate the performance of our proposed Thresholding Double Average Ratio procedure which utilizes the property of average and ratio for identifying the locations and the number of change points, we mainly run the simulations with the other competing procedures that have also been considered in \citet{ref4}, where we use the suggested sSIC-method for wild binary segmentation. Precisely, the methods are CumSeg as described in \citet{ref10} , SMUCE as described in \citet{ref6}, as well as WBS as considered in \citet{ref4}. For both standard change point scenario and local alternatives situation, the implementation of detection requires the choice of paramters, $\alpha_n$, $\tau$ and $c_n$. In order to avoid getting trouble in choosing parameter, we are to propose a data-driven algorithm for automatically selecting some parameters based on data. The specific algorithm have been given in Algorithm \ref{algo}. Another important thing in practice is to choose an appropriate thresholding. Threshold value will not be decided by data as we can give a expanation in theory. Theoretically, threshold should be $0$ becasue our statistics drop from 1 to $0$ at change point. However, in our simulation and real data, there is always some error that we can't avoid. Thus, we will relax the strict requirement and take it as a larger value. Here we use 0.5 as our choice. We have used normal distribution with mean changing from 1 to 4. As could be seen from Picture \ref{pic1}, the statistics drop below 0.5 clearly and thus the change points is the index attaching the minimum points.

\begin{algorithm}[!htbp]\label{algo}
	\caption{How to estimate change points}
	\KwIn{$ X \in \mathcal R^{n\times 1}$}
	Take $\alpha_n=n^{0.6}/3$, $\tau=0.5$,  $c_n=\sqrt{\log n\over \alpha_n}$and then perform the Double Average Ratio steps directly to acquire a preliminary estimation $\hat z_i,$ for $i=1,...,k. $\;
	
	Calculate the variance for each segementation $\hat \sigma_i, i=1,...,k,$ and average them, $\bar \sigma={{1\over k}\sum_{i=1}^k \sigma_i}.$\;
	Take $c_n=\sqrt{\log \over \alpha_n} * {1\over \bar \sigma} $, and perform TMAR construction steps again. \;
	Estimate $\hat z_i$ based on parameters we select in step 3.\;
	\KwOut{$\hat z_i$}
\end{algorithm}

For each example, 1000 replications is used to approximate the distribution of $\hat K-K$, where $\hat K_n$ is the estimated number of change points.

\subsection{Mean}

\noindent \emph{Part 1: CP model}

CP Model: Both the number and locations of change points are fixed. We adopt the blocks setting which is wildly used in the literature (\citet{ref4}). Specifically, $K_n=11$ and change points at 161, 323, 485, 638, 801, 967, 1132, 1299, 1465, 1632, 1794, and the values between 1, 3, 2, -1, 1, 3, 2, 5, 1, -2, 3, 0.

Detecting mean shift in a univariate observations has been widely discussed in the literature. In this section, two different change point model will be both considered. Four scenarios of error distribution will also be taken into account:
\begin{itemize}
	\item  (i) $\varepsilon_i \backsim^{iid} N(0,1)$,
	\item  (ii)$\varepsilon_i \backsim^{iid} N(0,3)$,
	\item   (iii) $\varepsilon_i \backsim^{iid} 7*Uniform(-1,1)$,
	\item (iv)$\varepsilon_i \backsim^{iid} 3*t_3$,
\end{itemize}
where $t_v$is the Student's t-distribution with degree of freedom $v$. The model we simulate could be represented as the following form:
\begin{equation}
X_i=\mu_i+\varepsilon_i.
\end{equation}
There are two models we will take into consideration. The first one is standard change point model which is usually used in the literature. While the second one is CP alternative model, which considers the weak magnitude signal for the change.

In this part, standard change point model which is used in the literature is considered, which means both the number and locations of change point are fixed, where we set $n=2048$. The signal function $\mu_i$s are chosen as a piecewise constant function with $K=11$. Table \ref{t4} compares the performance of TMAR and other competing procedures under different error settings. All the algorithms were run on the same data matrices and the  the distribution of $\hat K-K$ is reported.

It can be seen from table \ref{t4} that TMAR has a competitive performance for the change point estimation task. When $\sigma=1$, the other methods would perfoamce better than us. However, when $\sigma^2=3$, all the other methods would underestimate the number of change points. Though our method would also perform not so good, it is better than the others. When error type change from normal to some other distribution, our method perform obviously better than the others. When the error follows a uniform distribution, other methods would also underestimate change points number, but TMAR perform more stable compared with them. For $t$-distribution, cumSeg seriously underestimated change points while WBS and SMUCE would overestimate. They all doesn't work at all. Our method in some extent over estimate, but perform more stable than the others. Actually it is easy to understand. As in our setting the magnitude between change points are small, all the methods might perform not as good as we image. However, our method would still perform more stable than the others.

\noindent \emph{Part 2: CP local alternative model}

CP Alternative Model: In this model, we take magnitude smaller. Specifically, $K=11$ and change points at 161, 323, 485, 638, 801, 967, 1132, 1299, 1465, 1632, 1794, and the values between 0, 0.7, 0, -0.7, 0.7, 0 , 2, 2.7, 0, -2.7, -2, 0.
In this section, we are considering the following four types of error distribution.
\begin{itemize}
	\item  (i) $\varepsilon_i \backsim^{iid} N(0,1)$,
	\item  (ii)$\varepsilon_i \backsim^{iid} N(0,3)$,
	\item   (iii) $\varepsilon_i \backsim^{iid} 7*Uniform(-1,1)$,
	\item (iv)$\varepsilon_i \backsim^{iid} t_3$,
\end{itemize}

We are interested in the performance of our method and the others when facing the scenario of weak signal. In this situation, we have also presented the result of multiple change point detection. The distribution of $\hat K-K$ have been reported in Table \ref{t3}, displaying the error between the truth and the estimated location.

From table \ref{t3}, we can observe that our method perform obviously better than the other methods. Under the local alternative model, cumSeg, WBS and SMUCE could not estimate accurately, and even they couldn't detect the existence of change point. However, our approach can still work very well in terms of the number of change point. When the error type is normal distribution with $\sigma^2=3$, TMAR perform not as good as we image, but still more accurate than the other competitiors. It can be seen that our procedure is more robust than the other procedures from an overall view. Actually this result is easy to understand. all the other approaces have not considered the local alternatives. They all assumed that the signal is strong enough. It is not always that case. When the signal is weak, they could not work at all.

We now suggest an iterative algorithm to detect change points, which could be used in the short spacing scenarios. Note that in our method, the segment length must be larger than $3\alpha_n$ where $\alpha_n= O(n^{0.6})$. When the spacings between two change points is short, the length may be larger than the spacing and the change point within the segment cannot be found out. To attenuate the severity of this difficulty, the iterative algorithm is as follows. 
\begin{algorithm}[!htbp]\label{algo2}
	\caption{How to estimate change points for short spacing}
	
	\KwIn{$ X \in \mathcal R^{n\times 1}$}
	
	Start with the original data as the initial segement.\;
	
	Perform Algorithm \ref{algo} on the segementations and get an estimation of change points for each segementation.\;
	
	For each segementation in the last step perform Algorithm \ref{algo} untill there is no point of $T_n(i)$ found lower than $0.5$.\;
	
	\KwOut{All $\hat z_i$}
\end{algorithm}

\subsection{Variance}

The idea of detecting changes in mean can be easily extended to the variance change point problem. Similarly, four scenarios of error distribution will also be taken into account in this part:
\begin{itemize}
	\item  (i) $\varepsilon_i \backsim^{iid} N(0,1)$,
	\item  (ii)$\varepsilon_i \backsim^{iid} N(0,3)$,
	\item   (iii) $\varepsilon_i \backsim^{iid} 7*Uniform(-1,1)$,
	\item (iv)$\varepsilon_i \backsim^{iid} 3*t_3$,
\end{itemize}
where $t_v$is the Student's t-distribution with degree of freedom $v$. However, the model we simulate is different from the mean, which could be represented as the following form:

We will take the model:
\begin{equation}
X_i=\sigma_i \varepsilon_i.
\end{equation}
Similar to the last part, we will compare the performance of number of change points of four methods including SUMCE, BS which was introduced in last part and PELT which was considered in \citet{ref25}. In this part, standard change point model which is used in the literature is considered, which means both the number and locations of change point are fixed, where we set $n=2048$. The signal function $\mu_i$s are chosen as a piecewise constant function with $K=11$. Table \ref{t2} compares the performance of TMAR and other competing procedures under different error settings. All the algorithms were run on the same data matrices. We report the distribution of $\hat K-K$.

Table \ref{t2} gives the results of different error settings. Again we find that the performance of TMAR is very encouraging on all performance measures. For all situations, SMUCE and BS doesn't work at all for variance change point situation. Under normal distribution with $\sigma=1$, PELT is more accurate than our procedures however, when $\sigma^2=3$ our approach is more precise than PELT. When the error typr change to uniform distribution and t-distribution, our method performs more robust than PELT. All in all, our approach could deal with variance change point problem more precisely and robust.

\begin{figure}
	\centering
	\includegraphics[width=4.5in]{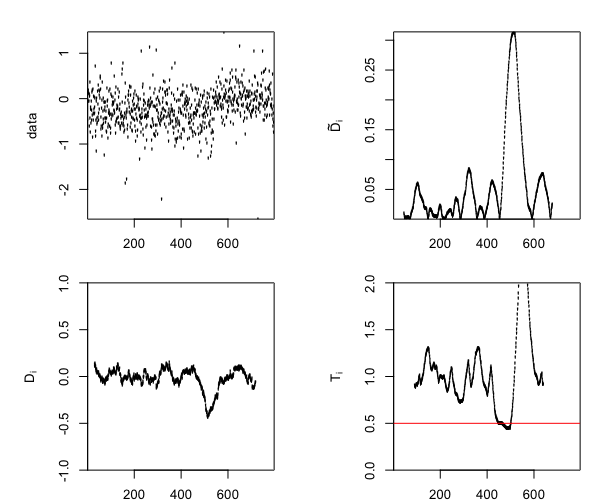}
	\caption{The plots for the Array CGH data}
	\label{picreal}
\end{figure}

\begin{figure}
	\centering
	\includegraphics[width=5.5in]{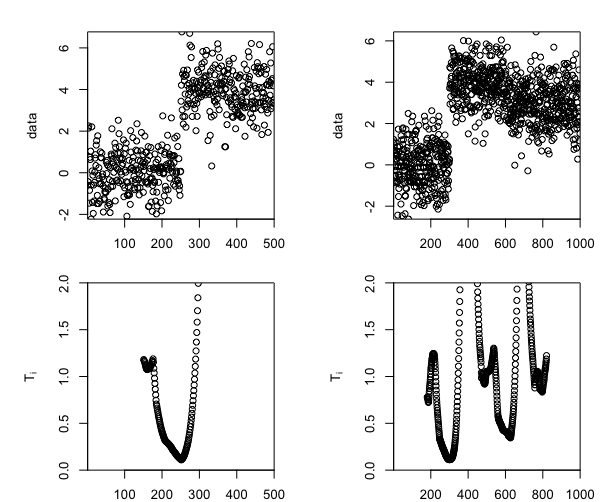}
	\caption{The upper two plots are for scatter points and the lower two plots are for the criterion functions $T_n(\cdot)$ for mean changes of normal distributions at the change points. }
	\label{pic1}
\end{figure}

\csection{Real Data Examples}
Consider  an Array CGH data set, which shows aberrations in genomic DNA. The observations are normalized glioblastoma profiles from the data set of \citet{ref_real}. We now detect regions on which the observations jump from $0$. Compute $T_n(i)$ about the array CGH profile of chromosome 13 in GBM31.
The threshold  $\tau=0.5$ and ridge $c_n$ are selected as before.
In Figure~\ref{picreal}, we plot the original data, $D_n(i)$, $\tilde D_n(i)$ and $T_n(i)$. From $D_n(i)$, we may see that the magnitudes of changes are small except for a point between $500$ and $600$, which is also smaller than $0.5$. $\tilde D_n(i)$ also shows this pattern, but the point is more stood out although the magnitude is still small about  $0.4$.  The plot of $T_n(i)$ presents a curve to clearly indicate that this point can be regarded as a change point.
$T_n(i)$ also suggests the location of the change point is the number $579$. From all four plots, the determination of this location would be reliable.

\csection{Discussion}
This paper has developed a new method to estimate the change point location. We have propoesed several setps procedure to eatimate the location and the number of change point at the same time. Our approach is easy to implement and more direct. In addition, our approach could be extended to a more general case besides mean and variance case which was considered in this paper. For example, our approach might be used to detect change point in distribution by the information of the difference between empirical distribution function. But it is difficult to deal with a sequence of the functions. Besides, our approach might also be applied to multivariate data which was considered in \citet{ref5}. However a reasonable measurement is not easy to find. \citet{ref3} clarify a high dimensional change point detection method. Under some conditions, our method might also be used to deal with change point detection problem in high dimensional data. But when coping with high dimensional data it is difficult to solve with caculation problem. More general, when coping with change point problem in functional data mentioned in \citet{functional}, our approach might also work but perhaps, some other conditions need to be added.

As we commented before, this method has the limitation to handle the short spacing issue. As an attempt, we once considered an iterative algorithm  to  partly alleviate this determination difficulty. The basic idea is that after performing the algorithm, the locations of some change points have been identified and estimated, we then perform the algorithm again with the samples between any two consecutive locations to see whether there are other change points. In such a procedure, the sample sizes  could be very much reduced, the segment lengths could then be much shorter. A small numerical study showed that some more locations are regarded as change points. However, as the length is shorter, the corresponding objective function  became oscillating such that the algorithm could produce an over-estimation. The numerical study showed that this phenomenon became more obvious with larger variance. Therefore, how to combine this method with existing  methods that can handle short spacing issue is an interesting topic in a further study.

\newpage
\csection{Appendix}
In this section, we present the proofs of the theoretical results.
\renewcommand{\theequation}{A.\arabic{equation}}
\setcounter{equation}{0}
\subsection{Appendix A: Two lemmas}
We give two lemmas first.
\begin{lemma}\label{lemma1}
	Assume that $X_i-EX_i$ are independent identically distributed random variables and ${ n^{1/4} \log n\over {\sqrt\alpha_n}}\rightarrow 0$,
	\begin{equation}
	\Pr\{\max_{1\leq i\leq n}\Big||\tilde D_n(i)|-|D(i)| \Big| >\tau_n \}=o(1)
	\end{equation}
	where $\tau_n=O(\sqrt{\log n\over \alpha_n})$.
\end{lemma}

\textbf{Proof of Lemma \ref{lemma1}}
We first rewrite $\tilde D_n(i)$ as a sum of independent variables:
\begin{eqnarray}
\tilde D_n(i)&=&{1\over \alpha_n^2}\bigg\{\sum_{j=i}^{i+\alpha_n-1}(j-i+1)X_j+\sum_{i+\alpha_n}^{i+2\alpha_n-1}(3\alpha_n-2j+2i-2)X_j \nonumber\\
&&+\sum_{i+2\alpha_n}^{i+3\alpha_n-1}(3\alpha_n-j+i-1)X_j\bigg\}.
\end{eqnarray}
Then  the variance  of $\tilde D_n(i)$ equals, for a constant $C>0$:
\begin{eqnarray}
&&Var\bigg\{ {1\over\alpha_n^2}(\sum_{j=i}^{i+\alpha_n-1}(j-i+1)X_j+\sum_{j=i+\alpha_n}^{i+2\alpha_n-2}(2i+3\alpha_n-2j-2)X_j \nonumber\\
&&+\sum_{i+2\alpha_n-1}^{i+3\alpha_n-2}(i+3\alpha_n-j-1)X_j)  \bigg \}
\nonumber\\&=&{Var(X_1)\over \alpha_n^4}  (\sum_{i=1}^{\alpha_n} 2*i^2+\sum_{h=1}^{\alpha_n}(3\alpha_n-2h)^2) := {C^2\over \alpha_n}=\sigma_n^2.
\end{eqnarray}
It is obvious that the variance of $\tilde D(i)$ is then free of the index $i$ with $\sigma_n=C/\sqrt\alpha_n$. In addition, as $\tilde D_n(i)$ is a weighted sum of $\{X_i\}_{i=1}^n$, we then further rewrite it. Define a  weight function $w_n(t,j)$ as denoting $[nt]$ as the largest integer that is smaller or equal to $[nt]$,
\begin{equation*}
\begin{aligned}
w_n(t,j)&=\operatorname{I}\{[nt]\le j\le [nt]+\alpha_n-1 \}{(j-[nt]+1)\over \alpha_n^2}
\\&\quad +\operatorname{I}\{[nt]+\alpha_n\le j\le [nt]+2\alpha_n-1 \}{(3\alpha_n-2j+2[nt]-2)\over \alpha_n^2}
\\&\quad +\operatorname{I}\{[nt]+2\alpha_n\le j\le [nt]+3\alpha_n-1 \}{(3\alpha_n-j+[nt]-1)\over \alpha_n^2},
\end{aligned}
\end{equation*}
where $\operatorname{I}\{B\}$ denotes indicator function of set $B$. As for evert $i$ there exists $t_i\in (0,1)$ such that $i=[nt_i]$, we have
\begin{equation}
\begin{aligned}
w_n(t_i,j)&=\operatorname{I}\{[nt_i]\le j\le [nt_{i}]+\alpha_n-1 \}{(j-i+1)\over \alpha_n^2}
\\&\quad +\operatorname{I}\{[nt_{i}] +\alpha_n\le j\le [nt_{i}]+2\alpha_n-1 \}{(3\alpha_n-2j+2[nt_{i}]-2)\over \alpha_n^2}
\\&\quad +\operatorname{I}\{[nt_{i}]+2\alpha_n\le j\le [nt_{i}]+3\alpha_n-1 \}{(3\alpha_n-j+[nt_{i}]-1)\over \alpha_n^2}.
\end{aligned}
\end{equation}
$\tilde D_n(i)$  can then be rewritten as $\tilde D_n(i)=\sum_{j=1}^nw_n(t_i,j)X_j$. Then $\tilde D_n(i)-\tilde D(i)=\sum_{j=1}^nw_n(t_i,j)(X_j-E(X_j))$.
Thus we have
\begin{equation*}
\frac{\tilde D_n(i)-\tilde D(i)} {\sigma_n}=\sum_{j=1}^n{w_n(t_i,j)\over \sigma_n}(X_j-E(X_j))
\end{equation*}
Let $\tilde w_n(t_i,j)={w_n(t_i,j)\over \sigma_n}$, $Y_n(t_i)={\tilde D_n(i)-\tilde D(i)/ \sigma_n}$ and $e_j=X_j-E(X_j)$.
Then we have that
\begin{equation}
Y_n(t_i)=\sum_{j=1}^n \tilde w_n(t_i,j)e_j,
\end{equation}
where  $\tilde w_n(t_i,j)$ can be seen as a special case of Equation (18) in \citet{ref2}.
In addition, define $\Omega_n(t_i)=|\tilde w_n(t_i,1)|+\sum_{j=2}^n|\tilde w_n(t_i,j)-\tilde w_n(t_i,j-1)|$ and $\Omega_n=\max_{1\leq i\leq n}\{\Omega_n(t_i) \}$. Some elementary calculations lead to
\begin{equation}\label{omega}
\Omega_n(t_i)={4\alpha_n+3\over \alpha_n^2\sigma_n}.
\end{equation}
As $\Omega_n(t_i)$ is free of $i$ and then    $\Omega_n={4\alpha_n+3\over \alpha_n^2\sigma_n}$.
The application of Theorem 3 in \citet{wuaop} and Equation (6)  in \citet{ref2} yield that there exists a Gaussian process below with the standard Brownian motion $\mathbb B(\cdot)$,
\begin{equation}\label{ynstar}
Y_n^*(t_i)=\sum_{j=1}^n\tilde w_n(t_i,j)\sqrt {Var(X_1)}\{\mathbb B(j)-\mathbb B(j-1) \}
\end{equation}
such that almost surely for all $i$
\begin{equation}
|Y_n(t_i)-Y_n^*(t_i)|\leq o({\Omega_n(t_i)n^{1/4}\log n }),
\end{equation}
and then
\begin{equation}
\max_{1\leq i\leq n} |Y_n(t_i)-Y_n^*(t_i)|= o({ \Omega_nn^{1/4}\log n }).
\end{equation}
This yields that almost surely
\begin{equation}
\begin{aligned}
\max_{1\leq i\leq n} |Y_n(t_i)|&=\max_{1\leq i\leq n} |Y_n(t_i)-Y_n^*(t_i)+Y_n^*(t_i)|
\\&\leq \max_{1\leq i\leq n} |Y_n^*(t_i)|+\max_{1\leq i\leq n}|Y_n(t_i)-Y_n^*(t_i)|
\\&\leq  \max_{1\leq i\leq n} |Y_n^*(t_i)|+o(\Omega_nn^{1/4}\log n ),
\end{aligned}
\end{equation}
and
\begin{equation}
\begin{aligned}
\max_{1\leq i\leq n}\Big||\tilde D_n(i)|-|\tilde D(i)| \Big|/\sigma_n &
\leq \max_{1\leq i\leq n}\Big| \tilde D_n(i)-\tilde D(i)\Big|/\sigma_n
\\&=\max_{1\leq i\leq n}\Big| Y_n(t_i) \Big|
\\&\leq \max_{1\leq i\leq n} |Y_n^*(t_i)|+o(\Omega_nn^{1/4}\log n ).
\end{aligned}
\end{equation}

Due to the fact $\sigma_n=O(1/\sqrt\alpha_n)$ and the result in (\ref{omega}), we can see that $\Omega_n={4\alpha_n+3\over \alpha_n^2\sigma_n}=O(1/\sqrt\alpha_n)$. By the condition ${n^{1/4}\log n\over \sqrt\alpha_n}\rightarrow 0$, we have for any $\tau_n$
\begin{equation}
\begin{aligned}
\Pr\{\max_{1\leq i\leq n}\Big||\tilde D_n(i)|-|\tilde D(i)| \Big| >\tau_n \}
&=\Pr\{\max_{1\leq i\leq n}\Big||\tilde D_n(i)|-|\tilde D(i)| \Big|/\sigma_n >\tau_n/\sigma_n \}
\\&\leq \Pr\{\max_{1\leq i\leq n}\Big|Y_n^*(t_i)\Big|+o({n^{1/4}\log n\over \sqrt\alpha_n}) >\tau_n/\sigma_n \}
\\&\leq \Pr\{\max_{1\leq i\leq n}\Big|Y_n^*(t_i)\Big|+1 >\tau_n/\sigma_n \}.
\end{aligned}
\end{equation}

From (\ref{ynstar}), we have
\begin{equation}
\operatorname{Var} (Y_n^*(i))={Var(X_1)\over \sigma_n^2\alpha_n^4}  (\sum_{i=1}^{\alpha_n} 2*i^2+\sum_{h=1}^{\alpha_n}(3\alpha_n-2h)^2)=1
\end{equation}
In other words, $Y_n^*(t_i)$ follows  the standard normal distribution, and thus with an application of Proposition 2.1.2 in \citet{roman2017}, we have, for large $\tau_n/\sigma_n$,
\begin{equation}\label{a14}
\begin{aligned}
\Pr\{\max_{1\leq i\leq n}\Big|Y_n^*(t_i)\Big|+1>\tau_n/\sigma_n \}\leq& n\max_{1\leq i\leq n}\Pr\{\Big|Y_n^*(t_i)\Big|+1 >\tau_n/\sigma_n \}
\\&= n\Pr\{\Big|Y_n^*(t_1)\Big|>\tau_n/\sigma_n-1 \}
\\&\leq n/({\tau_n\over \sigma_n}-1)\exp\{ -{1\over 2}({\tau_n\over \sigma_n}-1)^2 \}.
\end{aligned}
\end{equation}
Taking $\tau_n/\sigma_n={\sqrt{2\log n}}+ 1$, we have as $n\rightarrow \infty$
\begin{equation*}
\begin{aligned}
n/({\tau_n\over \sigma_n}-1)\exp \left\{ -{1\over 2}({\tau_n\over \sigma_n}-1)^2 \right\}
&=\exp\left\{ \log n -\log {\sqrt{2\log n}}  -\log n \right \}
\\&=\sqrt{1\over 2\log n} \rightarrow 0.
\end{aligned}
\end{equation*}
 That is when $\tau_n=\sigma_n({\sqrt{2\log n}}+ 1)=O(\sqrt {\log \alpha_n}/ \sqrt {\alpha_n})$ and $n\rightarrow \infty$, we have
\begin{equation}
\Pr\{\max_{1\leq i\leq n}\Big||\tilde D_n(i)|-|\tilde D(i)| \Big| >\tau_n \}\leq \sqrt{1 \over 2\log n} \rightarrow 0.
\end{equation}
This means that  $\max_{1\leq i\leq n}\Big||\tilde D_n(i)|-|\tilde D(i)| \Big|=O_p(\sqrt{\log n\over \alpha_n}).$
We complete the proof of Lemma \ref{lemma1}.

For the consistency of the estimated change points defined in the criterion, we first give the detailed computation of $\tilde D(i)$. It is easy to see that
\begin{eqnarray*}
	|\tilde D(i)|=\left\{
	\begin{array}{lll}
0, &z_{k-1}+\alpha_n \leq i\leq z_k-2\alpha_n; \\

{1+\cdots+(i-(z_k-2\alpha_n)) \over \alpha_n^2}\beta_k, &z_k-2\alpha_n<i\leq z_k-\alpha_n; \\

{{[(i-(z_k-\alpha_n-1))+\cdots +\alpha_n]}+{[(\alpha_n-1)+\cdots +(\alpha_n-(i-(z_k-\alpha_n)))]}\over \alpha_n^2}\beta_k, & z_k-\alpha_n<i \leq z_k-{\alpha_n\over 2}; \\

{{[(z_k-i+2)+\cdots +\alpha_n]}+{[(\alpha_n-1)+\cdots +(\alpha_n-(z_k-i+1))]}\over \alpha_n^2}\beta_k, &z_k-{\alpha_n\over 2} < i \leq z_k; \\

{1+\cdots+((z_k+\alpha_n)-i+1) \over \alpha_n^2}\beta_k, &z_k< i \leq z_k+\alpha_n; \\
0, &z_k+\alpha_n<i\leq z_{k+1}-2\alpha_n.

	\end{array}
	\right.
\end{eqnarray*}
From this formula, we  have a more detailed calculation that will be used in the proof of Lemma~2 and Theorem~2.1:
\begin{eqnarray}\label{mean}
	|\tilde D(i)|=\left\{
	\begin{array}{llll}
		0, &z_{k-1}+\alpha_n \leq i\leq z_k-2\alpha_n; \\
		
		{(i-z_k+2\alpha_n+1)\cdot(i-z_k+2\alpha_n) \over \alpha_n^2}\beta_k, &z_k-2\alpha_n<i\leq z_k-\alpha_n; \\
		
		{-i^2-\alpha_ni+2iz_k-i+z_k-z_k^2+\alpha_nz_k+{1\over 2}(\alpha_n^2-\alpha_n)\over \alpha_n^2}\beta_k,
		& z_k-\alpha_n<i < z_k-{\alpha_n\over 2}-{B_n}; \\
		
		({3\over 4}-{B_n^2\over \alpha_n^2}+{B_n\over \alpha_n^2})\beta_k, &i=z_k-{\alpha_n\over 2}-{B_n}; \\
		
		{-i^2-\alpha_ni+2iz_k-i+z_k-z_k^2+\alpha_nz_k+{1\over 2}(\alpha_n^2-\alpha_n)\over \alpha_n^2}\beta_k,
		& z_k-{\alpha_n\over 2}-{B_n}<i < z_k-{\alpha_n\over 2}; \\
		
		{3\over 4}\beta_k, &i=z_k-{1\over 2}\alpha_n;  \\
		
		{i^2+\alpha_ni-2iz_k+i-z_k+z_k^2-\alpha_nz_k-{1\over 2}(\alpha_n^2-\alpha_n)\over \alpha_n^2}\beta_k, &z_k-{\alpha_n\over 2} < i < z_k-{\alpha_n\over 2}+{B_n}; \\
		
		({3\over 4}-{B_n^2\over \alpha_n^2}+{B_n\over \alpha_n^2})\beta_k, &i=z_k-{\alpha_n\over 2}+{B_n}; \\
		
		{i^2+\alpha_ni-2iz_k+i-z_k+z_k^2-\alpha_nz_k-{1\over 2}(\alpha_n^2-\alpha_n)\over \alpha_n^2}\beta_k, &z_k-{\alpha_n\over 2} +{B_n} < i \leq z_k; \\
		
		{(-i+z_k+\alpha_n+2)(-i+1+\alpha_n+z_k) \over \alpha_n^2}\beta_k, &z_k< i \leq z_k+\alpha_n; \\ 
				0, &z_k+\alpha_n<i\leq z_{k+1}-2\alpha_n.
	\end{array}
	\right.
\end{eqnarray}
We can then know that when $z_{k-1}-2\alpha_n\leq i\leq z_k-{\alpha_n\over 2}$, $|\tilde D(i)|$ monotonically  increases with $i$ while when $z_k-{\alpha_n\over 2}\leq i\leq z_k+\alpha_n$, $|\tilde D(i)|$ monotonically decreases.

Similarly, we can derive  $T(i)=$ ${|\tilde D(i)|+c_n}\over {|\tilde D(i+{3\alpha_n\over2})|+c_n}$  as:
\begin{eqnarray}\label{ti}
	T(i)= \left\{
	\begin{array}{llll}
		{0+c_n\over 0+c_n}=1,
		&z_{k-1}+\alpha_n\leq i \leq z_k-{7\over 2}\alpha_n, \\
		
		{0+c_n\over  	{(i-z_k+2\alpha_n+1)\cdot(i-z_k+2\alpha_n) \over \alpha_n^2}\beta_k+c_n},
		& z_k-{7\over 2}\alpha_n<i\leq z_k-{5\over 2}\alpha_n, \\

		{0+c_n\over { {-i^2-\alpha_ni+2iz_k-i+z_k-z_k^2+\alpha_nz_k+{1\over 2}(\alpha_n^2-\alpha_n)\over \alpha_n^2}\beta_k+c_n} } ,
		& z_k-{5\over 2}\alpha_n<i< z_k-{2\alpha_n}-B_n, \\

		{0+c_n\over { ({3\over 4}-{B_n^2\over \alpha_n^2}+{B_n\over \alpha_n^2})\beta_k+c_n} } ,
		& i= z_k-{2\alpha_n}-B_n, \\

		{0+c_n\over { {-i^2-\alpha_ni+2iz_k-i+z_k-z_k^2+\alpha_nz_k+{1\over 2}(\alpha_n^2-\alpha_n)\over \alpha_n^2}\beta_k+c_n} } ,
		& z_k-{2\alpha_n}-B_n<i< z_k-{2\alpha_n}, \\

		{0+c_n\over { {3\over 4}\beta_k+c_n} } ,
		& i= z_k-{2\alpha_n}, \\

		{
			{(i-z_k+{2}\alpha_n+1)\cdot(i-z_k+{2}\alpha_n) \over \alpha_n^2}\beta_k+c_n
			\over
			{i^2+\alpha_ni-2iz_k+i-z_k-z_k^2+z_k^2-\alpha_nz_k-{1\over 2}(\alpha_n^2-\alpha_n)\over \alpha_n^2}\beta_k+c_n
		},
		&z_k-{2\alpha_n}< i \leq z_k-{2}\alpha_n+B_n,\\

		{{B_n(B_n+1)\over \alpha_n^2}+c_n\over { ({3\over 4}-{B_n^2\over \alpha_n^2}+{B_n\over \alpha_n^2})\beta_k+c_n} } ,
		& i= z_k-{{2}\alpha_n}+B_n, \\

		{
			{(i-z_k+{1\over2}\alpha_n+1)\cdot(i-z_k+{1\over 2}\alpha_n) \over \alpha_n^2}\beta_k+c_n
			\over
			{i^2+\alpha_ni-2iz_k+i-z_k+z_k^2-\alpha_nz_k-{1\over 2}(\alpha_n^2-\alpha_n)\over \alpha_n^2}\beta_k+c_n
		},
		&z_k-{{2}\alpha_n}+B_n< i \leq z_k-{3\over 2}\alpha_n,\\

		{
			{(i-z_k+{1\over2}\alpha_n+1)\cdot(i-z_k+{1\over 2}\alpha_n) \over \alpha_n^2}\beta_k+c_n
			\over
			{(-i+z_k+\alpha_n+1)\cdot(-i+z_k+\alpha_n+2) \over \alpha_n^2}\beta_k+c_n
		},
		&z_k-{3\over 2}\alpha_n< i \leq z_k-{\alpha_n},\\
		
		{
			{i^2+\alpha_ni-2iz_k+i-z_j-\alpha_nz_k-{1\over 2}(\alpha_n^2-\alpha_n)\over \alpha_n^2}\beta_k+c_n
			\over
			{((z_k+\alpha_n)-i+1)((z_k+\alpha_n)-i+2)\beta_k \over \alpha_n^2}+c_n
		},
		&z_k-{\alpha_n}< i \leq z_k-{1\over 2}{\alpha_n},\\
		
		{{i^2+\alpha_ni-2iz_k+i-z_j-\alpha_nz_k-{1\over 2}(\alpha_n^2-\alpha_n)\over \alpha_n^2}\beta_k+c_n
			\over 0+c_n},
		&z_k-{1\over 2}\alpha_n<i\leq z_{k},\\
		
		{{((z_k+{3\over 2}\alpha_n)-i+1)((z_k+{3\over 2}\alpha_n)-i+2) \over \alpha_n^2}+c_n \over
			0+c_n},
		&z_{k}<i\leq z_{k}+\alpha_n,\\
		
		{0+c_n\over 0+c_n}=1,
		&z_{k}+\alpha_n< i \leq z_{k+1}-{7\over 2}\alpha_n.
		
	\end{array}
	\right.
\end{eqnarray}

We now give another lemma and its proof.

\begin{lemma}\label{lemma2}
	Assume that $X_i-EX_i$ are independent identically distributed random variables, we could define $A^{d}=\{i: T(i)<d \}$ and $A_n^{d}=\{i: T_n(i)\leq d \}$ for any $0<d<1$. We have 	for any $d_1, \, d_2$ and $d_3$ with $0<d_3<d_1<d_2<1$.
	\begin{equation}\label{set1}
	\Pr\{A_n^{d_1}\subseteq A^{d_2} \}\rightarrow 1 \quad 	\Pr\{A^{d_3}\subseteq A_n^{d_1} \}\rightarrow 1.
	\end{equation}
Further, for any $k=1, \ldots, K$
the intervals $(m_k, M_k)$ are disjoint and each contains only one local minimizer $z_k-3 \alpha_n/2$ of $T(i)$.
Further,  for any $d$  with $0<d<1$,
	\begin{equation}\label{unif}
		\max_{i\in A_n^{d}} |T_n(i)-T(i)|=o_p(1).
	\end{equation}
\end{lemma}

\textbf{Proof of Lemma \ref{lemma2}}
To prove this lemma, we first analyse the properties of   $T_n(i)={\tilde D_n(i)+c_n\over \tilde D_n(i+{3\over 2}\alpha_n)+c_n}$ around the point $z_k-2\alpha_n$ where $z_k$ is the change point. Write it as
\begin{equation}\label{Tn}
\begin{aligned}
T_n(i)&={|\tilde D_n(i)|+c_n\over |\tilde D_n(i+{3\over 2}\alpha_n)|+c_n}
\\&={|\tilde D_n(i)|-|\tilde D(i)|+|\tilde D(i)|+c_n\over |\tilde D_n(i)|-|\tilde D(i+{3\over 2}\alpha_n)|+|\tilde D(i+{3\over 2}\alpha_n)|+c_n}
\\&={O_p({\sqrt{\log n}\over \sqrt{\alpha_n}})+|\tilde D(i)|+c_n\over O_p({\sqrt{\log n}\over \sqrt{\alpha_n}})+|\tilde D(i+{3\over 2}\alpha_n)|+c_n}
\end{aligned}
\end{equation}
For the flat parts in the sequence with $|\tilde D(i)|=0$ for all $i$,  we have
\begin{equation}
T_n(i)={O_p({\sqrt{\log n}\over \sqrt{\alpha_n}})+0+c_n\over O_p({\sqrt{\log n}\over \sqrt{\alpha_n}})+0+c_n} = o_p(1).
\end{equation}
When a change point appears, we have that, from (\ref{mean}) and the discussion right below it, for $\forall i\in [z_k-{7\over 2}\alpha_n,z_k-2\alpha_n]$, $|\tilde D(i)|=0$, $|\tilde D(i+{3\over 2}\alpha_n)|$  monotonically increases and at $i=z_k-2\alpha_n$,  we have
\begin{equation}\label{Tn1}
\begin{aligned}
T_n(z_k-{2}\alpha_n)
={|\tilde D_n(z_k-2\alpha_n)|+c_n\over|\tilde D_n(z_k-{1\over 2}\alpha_n)|+c_n}
={O_p({\sqrt{\log n} \over \sqrt{\alpha_n}})+0+c_n \over  O_p({\sqrt{\log n} \over \sqrt{\alpha_n}})+{3\over 4}\beta_k+c_n}=o_p(1).
\end{aligned}
\end{equation}
As we discussed before, for any  $i \in [z_k-2\alpha_n,z_k-{1\over 2}\alpha_n]$, $|\tilde D(i)|$  monotonically increases, and $|\tilde D(i+{3\over 2}\alpha_n)|$ monotonically decreases, then $T_n(i)$ uniformly converges to the monotonically increasing $T(i)$ and
\begin{equation}\label{Tn2}
\begin{aligned}
T_n(z_k-{1\over 2}\alpha_n)
={|\tilde D_n(z_k-{1\over 2}\alpha_n)|+c_n\over|\tilde D_n(z_k+\alpha_n)|+c_n}
={O_p({\sqrt{\log n} \over \sqrt{\alpha_n}})+{3\over 4}\beta_k+c_n \over {O_p({\sqrt{\log n} \over \sqrt{\alpha_n}})+0+c_n } }\xrightarrow{P} \infty .
\end{aligned}
\end{equation}

{\bf Step 1} To prove the subset equations in (\ref{set1}) and the uniform convergence in (\ref{unif}). Define  $A^{d_2}=\{i: T(i)<d_2\}$ and $A_n^{d_1}=\{i: T_n(i)<d_1\}$ where $d_1<d_2$.
Recall  the decomposition of (\ref{Tn}). 
By the definition of $ A_n^{d_1}$, we have for all $i\in A_n^{d_1}$, we have $T_n(i)\leq d_1$. Then,
\begin{equation*}
o_p(c_n)+|\tilde D(i)|+c_n\leq d_1(o_p(c_n)+|\tilde D(i+{3\over 2}\alpha_n)|+c_n).
\end{equation*}
That is,
\begin{equation*}
\begin{aligned}
|\tilde D(i)|+c_n&\leq d_1(|\tilde D(i+{3\over 2}\alpha_n)|+c_n)+o_p(c_n).
\end{aligned}
\end{equation*}
We can get, uniformly over all $i$, in probability, for large $n$
\begin{equation}
\begin{aligned}
T(i)&={|\tilde D(i)|+c_n\over  |\tilde D(i+{3\over 2}\alpha_n)|+c_n}
\\&\leq d_1+o(1)< d_2.
\end{aligned}
\end{equation}
In other words, with a probability going to one, $A_n^{d_1}\subseteq A^{d_2}=\{i: T(i)<d_2\}$. We can similarly prove that with a probability tending to one, $A^{d_3}\subseteq A_n^{d_1}$ for $d_3$ with $d_3<d_1<1$.

{\bf Step 2}. To prove that for any $k=1, \ldots, K$
the intervals $(m_k, M_k)$ are disjoint and each contains only one local minimizer $z_k-2 \alpha_n$ of $T(i)$. Consider  a value $d$ with $d>0.5$. Let $\tilde m_k$ and $\tilde M_k$ satisfy the following conditions:
\begin{equation*}
T(\tilde m_{k}-1)\geq d, \quad
T( \tilde m_{k})< d,
\end{equation*}
\begin{equation*}
T(\tilde M_{k})< d, \quad
T(\tilde M_{k}+1)\geq d.
\end{equation*}
Denote the interval $(\tilde m_k, \tilde M_k)$. From the previous proof, we can easily derive that in probability, $( m_k,  M_k)\subseteq (\tilde m_k, \tilde M_k).$ Further, from the properties, we also know that all $(\tilde m_k, \tilde M_k)$ are contained in $A^d$ and disjoint, also each interval contains only one local minimizer $z_k-{2}\alpha_n$ of $T(i)$. When we choose a value $d$ with $0<d<0.5$ we can  derive that in probability, $(\tilde m_k, \tilde M_k)\subseteq ( m_k,  M_k).$ Similarly, we also know that all $(\tilde m_k, \tilde M_k)$ are contained in $A^d$ and disjoint, also each interval contains only one local minimizer $z_k-{2}\alpha_n$ of $T(i)$. These two properties imply that in probability $( m_k,  M_k)$ are contained in $A_n^{0.5}$ and disjoint, also each interval contains only one local minimizer $z_k-{2}\alpha_n$ of $T(i)$.

{\bf Step 3.} To prove the uniform convergence of $T_n(i)$ to $T(i)$ over the set $A_n^{d_1}$.
As in probability $ A_n^{d_1}\subseteq A^{d_2}$ such that $T(i)\le d_2<1$, we consider a large set to derive the uniform convergence. For any $i\in A^{d_2}$, we have, uniformly,
\begin{equation*}
\begin{aligned}
T_n(i)-T(i)
&={|\tilde D_n(i)|+c_n\over |\tilde D_n(i+{3\over 2}\alpha_n)|+c_n}-{|\tilde D(i)|+c_n\over |\tilde D(i+{3\over 2}\alpha_n)|+c_n}
\\&={|(D_n(i)|+c_n)(|\tilde D(i+{3\over 2}\alpha_n)|+c_n)-(|\tilde D(i)|+c_n)(|\tilde D_n(i+{3\over 2}\alpha_n)|+c_n) \over (|\tilde D_n(i+{3\over 2}\alpha_n)|+c_n)(|\tilde D(i+{3\over 2}\alpha_n)|+c_n) }
\\&=\{{[(|\tilde D_n(i)|-|\tilde D(i))||(\tilde D(i+{3\over 2}\alpha_n)|+c_n)]\over (|\tilde D_n(i+{3\over 2}\alpha_n)|+c_n)(|\tilde D(i+{3\over 2}\alpha_n)|+c_n)}
\\&-{[(|\tilde D_n(i+{3\over 2}\alpha_n)|-|\tilde D(i+{3\over 2}\alpha_n)|)(|\tilde D(i)|+c_n)] \over (|\tilde D_n(i+{3\over 2}\alpha_n)|+c_n)(|\tilde D(i+{3\over 2}\alpha_n)|+c_n) }\}
\\&=\{{[o_p(c_n)(|\tilde D(i+{3\over 2}\alpha_n)|+c_n)]-[o_p(c_n)(|\tilde D(i)|+c_n)] \over (|\tilde D_n(i+{3\over 2}\alpha_n)|+c_n)(|\tilde D(i+{3\over 2}\alpha_n)|+c_n)}
\}
\\&= {o_p(c_n)\over o_p(c_n)+(|\tilde D(i+\frac {3\alpha_n}2)|+c_n)}-o_p(c_n)T(i)
\\&= {o_p(c_n)\over c_n} - o_p(c_n)=o_p(1).
\end{aligned}
\end{equation*}
Thus $\max_{i\in A^{d_2}}|T_n(i)-T(i)|=o_p(1)$. The proof is finished.

\subsection{Appendix B: Proof of Theorem \ref{theo1}}
We consider the first part in the theorem.
By Lemma~2, in probability $z_k-2\alpha_n\in (\tilde m_k, \tilde M_k)\subseteq A^{d}$ implies that $z_k-2\alpha_n \in ( m_k, M_k) \subseteq A_n^{0.5} \subseteq A^{d}$.  Thus uniformly  over $1\le k\le K$ in probability, we have
\begin{equation}
\tilde m_k\leq z_k-2\alpha_n \le \tilde  M_k.
\end{equation}
At the population level with $T(i)$'s, by the uniqueness of $z_k-2\alpha_n$ in the interval $( m_k, M_k)$, searching for $z_k-2\alpha_n$ in $( m_k,  M_k)$ is equivalent to searching for $z_k-2\alpha_n$ in the non-random $( \tilde m_k,  \tilde M_k)$ in probability.

Write $\hat z_k-2\alpha_n$ as the local minimizer of $T_n(i)$'s in the interval $( m_k,  M_k)\subseteq (\tilde m_k, \tilde M_k)\subseteq A^{d}$. Recall that by Lemma \ref{lemma2} $\max_{i\in A^{d_2}} | T_n(i)-T(i)|=o_p(1).$
We can then work on  each interval $(m_k, M_k)$. For any $k$ with $1\leq k \leq K$, from (\ref{ti}),  $T(z_k-2\alpha_n)$ is the only local minimum and
by the definition of $\hat z_k-2\alpha_n$,  $T_n(i)\geq T_n(\hat z_k-2\alpha_n)$  in the interval in probability. From (\ref{Tn1}) and (\ref{Tn2}), we have that, as $ |\tilde D(z_k-2\alpha_n)|=0$,
\begin{equation}
|\tilde D_n(z_k-2\alpha_n)|=O_p(\sqrt{\log n}/\sqrt{\alpha_n})=o_p(c_n)
\end{equation}
and, as $|\tilde D(z_k-{1\over 2}\alpha_n)|\big |=3\beta_k/4$,
\begin{equation}
|\tilde D_n(z_k-{1\over 2}\alpha_n)|-3\beta_k/4= O_p(\sqrt{\log n}/\sqrt{\alpha_n})=o_p(c_n).
\end{equation}
Further, from the calculation of $T(i)$ before, we can see that letting $B_n=\alpha_n (\log \alpha_n)^{-1/5} $, for any  $j=O(B_n)$
\begin{equation}
|\tilde D(z_k-2\alpha_n\pm j)|=O(c_n).
\end{equation}

To prove that $\hat z_k/z_k- 1=o_p(1)$, we only need to  prove that $|\hat z_k-z_k|=O_p(B_n).$ To this end,  applying the strictly decreasing and increasing monotonicity of $T(i)$ on the two sides of $z_k-2\alpha_n$ respectively, and the uniform convergence of $T_n(i)$ to $T(i)$ in probability in the set $A_n^{0.5}$,  we only need to show that $T_n(z_k-2\alpha_n\pm B_n)-T_n(z_k-2\alpha_n)>0$ in probability. Consider $T_n(z_k-2\alpha_n-B_n)$ first. Note that
\begin{equation}
T_n(z_k-2\alpha_n-B_n)={0+c_n+o_p(c_n) \over { ({3\over 4}-{B_n^2\over \alpha_n^2}+{B_n\over \alpha_n^2})\beta_k+c_n+o_p(c_n)}}.
\end{equation}
 Let $b_{n1}=({B_n^2\over \alpha_n^2}-{B_n\over \alpha_n^2})\beta_k$. To simplify the notations, in the following  all derivations are in probability. We can derive that
\begin{equation}\label{left}
\begin{aligned}
\quad &T_n(z_k-2\alpha_n-B_n)-T_n(z_k-2\alpha_n)\\
&={c_n+O({{\sqrt{\log n}}\over \sqrt\alpha_n})\over O({1\over \sqrt\alpha_n})+c_n+{3\over 4}\beta_k-b_n   }-{c_n+O({{\sqrt{\log n}}\over \sqrt\alpha_n})\over O({1\over \sqrt\alpha_n})+c_n+{3\over 4}\beta_k  }
\\&:={c_n+a_{n2}\over \beta_{n2}-b_{n1}   }-{c_n+a_{n1}\over \beta_{n1} }
\\&={(a_{n2}+c_n)\beta_{n1}-(a_{n1}+c_n)(\beta_{n2}-b_{n1}) \over \beta_{n1}(\beta_{n2}-b_{n1})  }
\\&={(a_{n1}+c_n)(\beta_{n1}-\beta_{n2}) + (a_{n2}-a_{n1})\beta_{n1} + (a_{n1}+c_n)b_{n1}  \over  \beta_{n1}(\beta_{n2}-b_{n1})   }
\\&={(a_{n1}+c_n)O({{\sqrt{\log n}}\over \sqrt\alpha_n}) + O({{\sqrt{\log n}}\over \sqrt\alpha_n})\beta_{n1} + (a_{n1}+c_n)b_{n1}  \over  \beta_{n1}(\beta_{n2}-b_{n1})   }
\\&={((a_{n1}+c_n)b_{n1})[O({{\sqrt{\log n}}\over {b_{n1}\sqrt\alpha_n}})+ O({{\sqrt{\log n}}\over {(a_{n1}+c_n)b_{n1}\sqrt\alpha_n}})\beta_{n1} + 1 ]\over  \beta_{n1}(\beta_{n2}-b_{n1})   }.
\end{aligned}
\end{equation}
When $(a_{n1}+c_n)b_{n1}\sqrt\alpha_n/{\sqrt{\log n}} \to \infty$, and ${ {b_{n1}\sqrt\alpha_n}} /{\sqrt{\log n}}\to \infty$, we then have for large $n$, the value in the 	brackets is larger than a positive constant and then the numerator is positive as $c_n\sqrt\alpha_n/{\sqrt{\log n}} \rightarrow \infty$ and $c_n>0$ such that 
$a_{n1}+c_n=c_n(1+{a_{n1}\over c_n})
=c_n(1+O({{\sqrt{\log n}}\over \sqrt \alpha_n c_n}))>0$ and
 $(a_{n1}+c_n)b_{n1}>0$.  We then have  $T_n(z_k-2\alpha_n-B_n)-T_n(z_k-2\alpha_n)>0$ when $b_{n1}\cdot c_n\cdot {\sqrt\alpha_n}={B_n^2\over \alpha_n^2}\cdot c_n\cdot \sqrt\alpha_n /{\sqrt{\log n}} > {B_n^2\over \alpha_n^2}\cdot \sqrt {\log \alpha_n } \rightarrow \infty$.

For $i=z_k-2\alpha_n+B_n$, we have
\begin{equation}
T_n(z_k-2\alpha_n+B_n)={{B_n(B_n+1)\over \alpha_n^2}\beta_k+c_n+o_p(c_n)\over { ({3\over 4}-{B_n^2\over \alpha_n^2}+{B_n\over \alpha_n^2})\beta_k+c_n+o_p(c_n)} }.
\end{equation}
Let $b_{n2}={B_n(B_n+1)\over \alpha_n^2}\beta_k$. We similarly have, in probability,
\begin{equation}\label{right}
\begin{aligned}
&T_n(z_k-2\alpha_n+B_n)-T_n(z_k-2\alpha_n)\\
&={c_n+O({{\sqrt{\log n}}\over \sqrt\alpha_n}) +b_{n2}\over c_n+O({{\sqrt{\log n}}\over \sqrt\alpha_n})+{3\over 4}\beta_k -b_{n1}  } - {c_n+O({{\sqrt{\log n}}\over \sqrt\alpha_n})\over O({{\sqrt{\log n}}\over \sqrt\alpha_n})+c_n+{3\over 4}\beta_k  }
\\&=: {c_n+a_{n3} +b_{n2}\over \beta_{n3} -b_{n1}  } - {c_n+a_{n1}\over \beta_{n1} }
\\&= {(c_n+a_{n3} +b_{n2})\beta_{n1} - (a_{n1}+c_n)\beta_{n3} + (a_{n1}+c_n)b_{n1} \over \beta_{n1}(\beta_{n3} -b_{n1})}
\\&= {(a_{n3} - a_{n1})\beta_{n1} + (a_{n1}+c_n)(\beta_{n1}-\beta_{n3})  + (a_{n1}+c_n)b_{n1} + b_{n2}\beta_{n1} \over \beta_{n1}(\beta_{n3} -b_{n1})}
\\&\geq {O({{\sqrt{\log n}}\over \sqrt\alpha_n})\beta_{n1} + (a_{n1}+c_n)O({{\sqrt{\log n}}\over \sqrt\alpha_n}) + b_{n2}\beta_{n1} \over \beta_{n1}(\beta_{n3} -b_{n1})}
\\&= { b_{n2}  [O({{\sqrt{\log n}}\over \sqrt\alpha_n b_{n2}})\beta_{n1} + (a_{n1}+c_n)O({{\sqrt{\log n}}\over \sqrt\alpha_n b_{n2}})  + \beta_{n1} ] \over \beta_{n1}(\beta_{n3} -b_{n1})}
\end{aligned}
\end{equation}
The inequality  is due to  $(a_{n1}+c_n)b_{n1}>0$.
Thus as long as $b_{n2}\cdot \sqrt\alpha_n/{\sqrt{\log n}} >B_n^2\alpha_n^{-3/2}/{\sqrt{\log n}} \rightarrow \infty$, the first term in the brackets converges to zero. Note that $a_{n1}$ and $c_n$  both tend to zero. The second term converges to zero. As $\beta_{n1}=O({{\sqrt{\log n}}\over \sqrt\alpha_n})+c_n+{3\over 4}\beta_k$, in which $O({{\sqrt{\log n}}\over \sqrt\alpha_n})$ and $c_n$ go  to zero, $\beta_{n1}$ then tends to $\beta_k$ and thus $\beta_{n1}$ is larger than zero for  large $n$. Therefore, $(O({{\sqrt{\log n}}\over \sqrt\alpha_n b_{n2}})\beta_{n1} + (a_{n1}+c_n)O({{\sqrt{\log n}}\over \sqrt\alpha_n b_{n2}})  + \beta_{n1} )$ is greater than zero. The whole numerator and then the difference is larger than zero such that $T_n(z_k-2\alpha_n+B_n)-T_n(z_k-2\alpha_n)>0$. 
Altogether,
when $B_n^2\cdot c_n\cdot \alpha_n^{-{3\over 2}}/{\sqrt{\log n}} \rightarrow \infty$, then
\begin{equation}
T_n(z_k-2\alpha_n\pm B_n)-T_n(z_k-2\alpha_n)>0.
\end{equation}
As we argued before, $\hat z_k$ cannot be larger than $z_k\pm B_n$ in probability. Also, based on the definition in Lemma \ref{lemma2}, we can get that $(z_k-2\alpha_n - B_n, z_k-2\alpha_n+B_n)\subset A_n^{d_1}$. That is
\begin{equation*}
-B_n+z_k-2\alpha_n\leq \hat z_k-2\alpha_n\leq B_n+z_k-2\alpha_n.
\end{equation*}
As ${B_n\over \alpha_n}\rightarrow 0$
\begin{equation*}
\begin{aligned}
|{\hat z_k-z_k\over \alpha_n}|\leq {B_n\over \alpha_n}\rightarrow 0
\end{aligned}
\end{equation*}
in probablity. In  other words, for any $\epsilon >0$, we have the uniform convergence over all $k\le K$: as $n\to \infty$
\begin{equation}
\begin{aligned}
P(\max_{1\le k\le K}|{\hat z_k-z_k\over \alpha_n}|<\epsilon )\to 1\end{aligned}
\end{equation}
This proves that uniformly over  all $k\le K$, $\hat z_k$ is a consistent estimator of $z_k$ in the above sense.
The proof of the first part of \ref{theo1} is finished.

 We now prove the second Part of Theorem \ref{theo1}.
From the  proof of the first part, we can see that we can consistently estimate all $z_k$ for $1\le k\le K$. Thus, clearly $\hat K=K$ with a probability going to one.

Now we prove the  third part of Theorem \ref{theo1}.
In the case with divergent $K$,   along with the steps in the proof of Lemma~\ref{lemma2} and of the first part of the theorem, we still have that $\max_k T_n(z_k-{2}\alpha_n)\to  0$ in probability.  That is, the local minima of $ T_n(z_k-{2}\alpha_n)$ can also converge to zero. The consistency can be proved almost the same as that for given $K$. Also $\hat K=K$ with a probability going to one in the divergent case.  We then omit the details and finish the proof.

\subsection{Appendix C: Proof of Theorem \ref{theo2}}

Denote the minimum change magnitude as $\beta_z=\min_{1\leq k \leq K_n}\beta_k$. $\beta_z$  converges to $0$ at the rate of $O((\log\alpha_n)^{-1/5})$ by the assumption.

From the  proof of Lemma~\ref{lemma2} and (\ref{ti}), we have that, letting $B_n=\alpha_n(\log \alpha_n)^{-1/10}$,  for any $j=O(B_n)$,
\begin{equation}
|\tilde D(z_k-2\alpha_n\pm j)|=O(c_n).
\end{equation}
To this end,  applying the strict monotonicity of $T(i)$, respectively, on the two sides of $z-2\alpha_n$, and the uniform convergence of $T_n(i)$ to $T(i)$ in probability in the set $A_n^{0.5}$,  we only need to show that $T_n(z_k-2\alpha_n\pm B_n)-T_n(z_k-2\alpha_n)>0$ in probability. In other words, we only need to check, similarly as those in (\ref{left}) and (\ref{right}),
\begin{equation}
b_{n1}\cdot c_n\cdot {\sqrt\alpha_n}/{\sqrt{\log n}}\rightarrow \infty
\end{equation}
where $b_{n1}=({B_n^2\over \alpha^2_n}-{B_n\over \alpha_n^2})\beta_z$.
As $\beta_z=O((\log\alpha_n)^{-1/5})$ and $B_n=\alpha_n(\log \alpha_n)^{-1/10}$, we have the above convergence. 
Then 
\begin{equation}
T_n(z_k-2\alpha_n\pm B_n)-T_n(z-2\alpha_n)>0.
\end{equation}
Thus $z_k- B_n\le \hat z_k \le z_k+ B_n $ in probability. As ${B_n\over \alpha_n}\rightarrow 0$, we have uniformly over all $k\le K$ in probability
\begin{equation*}
|{\hat z_k-z_k\over \alpha_n}|\leq {B_n\over \alpha_n}\rightarrow 0.
\end{equation*}
The proof is finished.

\subsection{Appendix D: Proof of Theorem \ref{theo3}}
We now prove the consistency of the estimators of the variance change points. From the criterion construction, the proof is very much similar to that for Theorem~\ref{theo1} as long as we pay attention to the rate of uniform convergence of $D_n(i)$ that is in this case the variance difference. Rather than only considering the first and second moment, we should take both second and forth moment into account. For both of the mean and variance scenario, the number of variables that $D_n(i)$ involves is the same. We then finish the proof without repeating the details that are used to prove Theorem~\ref{theo1}.

\newpage
\begin{singlespace}
	\scsection{REFERENCES}
	\begin{description}
		
		\newcommand{\enquote}[1]{``#1''}
		\expandafter\ifx\csname
		natexlab\endcsname\relax\def\natexlab#1{#1}\fi
		
		
		
		\bibitem[Bauer and Hackl, 1980]{ref21}
		Bauer, P. and Hackl, P. (1980) \enquote{An extension of the MOSUM technique for quality control} \textit{Technometrics}, \textbf{22}, 1-7

		\bibitem[Berkes et al., 2009]{functional}
		Berkes, I., Gabrys, R. Horvath, L. and Kokoszka, P. (2009) \enquote{Detecting changes in the mean of functional observations} \textit{Journal of the Royal Statistical Society: Series B}, \textbf{71}, 927-946
		

		\bibitem[Bredel et al., 2005]{ref_real}
		Bredel, M., Bredel, C., Juric, D., Harsh, G., Vogel, H., Recht, L. and Sikic, B. (2005) \enquote{High-resolution genome-wide mapping of genetic alterations in human glial brain tumors} \textit{Cancer Research}, \textbf{65}, 4088-4096
		
		\bibitem[Cao and Wu, 2015]{ref1}
		Cao, H. and Wu, W. B. (2015) \enquote{Changepoint estimation: another look at multiple testing problem} \textit{ Biometrika}, \textbf{102}, 974-980
		
		\bibitem[Chu et al., 1995]{ref22}
		Chu,C., Hornik, K. and Kuan, C.(1995) \enquote{MOSUM tests for parameter constancy} \textit{Biometrika}, \textbf{82}, 603-617
		
		
		\bibitem[Frick et al., 2014]{ref6}
		Frick, K., Munk, A. and Sieling, H. (2014) \enquote{Multiscale change point inference} \textit{Journal of the Royal Statistical Society: Series B}, \textbf{73}, 495-580
		
		\bibitem[Fryzlewicz, 2014]{ref4}
		Fryzlewicz, P. (2014) \enquote{Wild binary segmentation for multiple change-point detection} \textit{Annals of Statistics}, \textbf{42}, 2243-2281
		
		\bibitem[Gao et al., 2018]{ref7}
		Gao, Z., Shang, Z., Du, P. and Robertson, J. L. (2018) \enquote{Variance change point detection under a smootly-changing mean trend with application to liver procureent} \textit{Journal of American Statistical Association}, \textbf{114}, 1-9.
		
		\bibitem[Harchaoui and Levy-Leduc, 2010]{ref18}
		Harchaoui, Z. and Levy-Leduc, C.(2010).\enquote{Multiple change-point estimation with a total variation penalty} \textit{Journal of American Statistical Association}, \textbf{105}, 1480-1493
		
		
		\bibitem[Killick et al., 2012]{ref25}
		Killick, R., Fearnhead, P. and Eckley, I.A. (2012) \enquote{Optimal detection of changepoints with a linear computational cost} \textit{Journal of American Statistical Association}, \textbf{107}, 1590-1598
		
		

		
		\bibitem[Matteson and James, 2014]{ref5}
		Matteson, D. S. and James, N. A. (2014) \enquote{A Nonparametric Approach for Multiple Change Point Analysis of Multivariate Data} \textit{Journal of American Statistical Association}, \textbf{109}, 334-345
		

		\bibitem[Muggeo and Adelfio, 2011]{ref10}
		Muggeo, V. M. and Adelfio, G. (2011) \enquote{Sufficient Change Point Detection for Genomic Sequences of Continuous Measurements} \textit{Bioinformatics}, \textbf{27}, 161-166
		
		\bibitem[Page, 1954]{ref19}
		Page, E. S. (1954) \enquote{Continuous inspection schemes} \textit{Biometrika}, \textbf{41}, 110-115

		\bibitem[Preuss et al, 2015]{mocum}
		Preuss, P., Puchstein, R. and Dette, H. (2015) \enquote{Detection of multiple structural breaks in multivariate time series} \textit{Journal of American Statistical Association}, \textbf{110}, 654–668
		
		
		\bibitem[Pollak, 1987]{pollak}
		Pollak, M. (1987) \enquote{Average run lengths of optimal methods of detecting a change in distribution} \textit{Annals of Statistics}, \textbf{15}, 749-779
		
		\bibitem[Qu and Perron, 2007]{refred72}
		Qu, Z.  and Perron.  P. (2007) \enquote{ Estimating and testing structural changes in multivariate regressions} \textit{Econometrica}, \textbf{75}, 459-502
		
		\bibitem[Roman, 2017]{roman2017}
		Roman, V. (2007) \enquote{ High dimensional probability: an introduction with applications in data science} \textit{University of Michigen}



		
		\bibitem[Vostrikova, 1981]{ref20}
		Vostrikova, L. J. (1981) \enquote{Detecting disorder in multidimensional random processes} \textit{Soviet Mathematics Doklady}, \textbf{24}, 55-59
		
		\bibitem[Wang and Samworth, 2018]{ref3}
		Wang, T. and Samworth, R.J. (2018) \enquote{High dimensional change point estimation via sparse projection} \textit{Journal of the Royal Statistical Society: Series B}, \textbf{80}, 57-83
		
		\bibitem[Wu, 2007]{wuaop}
		Wu, W. B. (2007) \enquote{Strong invariance principles for dependent random varibles} \textit{The Annals of Probability}, \textbf{35}, 2294-2320
		
		\bibitem[Wu and Zhao, 2007]{ref2}
		Wu, W. B. and Zhao, Z. (2007) \enquote{Inference of trends in time series} \textit{Journal of the Royal Statistical Society: Series B}, \textbf{69}, 391-410
		
		\bibitem[Xia et al., 2015]{ref30}
		Xia, Q., Xu, W. and Zhu, L. (2015) \enquote{ Consistently determining the number of factors in multivariate volatility modelling} \textit{Statistica Sinica}, \textbf{25}, 1025-1044

		\bibitem[Yao, 1988]{ref16}
		Yao, Y, C.(1988) \enquote{Estimating the number of change points via Schwarz' criterion} \textit{Statistics and Probability Letters}, \textbf{6}, 181-189
		
		\bibitem[Yao and Au, 1989]{ref17}
		Yao, Y, C. and Au, S.T. (1989) \enquote{Least-squares estimation of a step function.} \textit{Sankhya Series A}, \textbf{51}, 370-381
		
		\bibitem[Zou et al., 2019]{zounum}
		Zou, C. Wang, G. and Li, R. (2019) \enquote{Consistent selection of the number of change-points via sample-splitting.} \textit{Annals of Statistics} To appear
		
		\bibitem[Zhu et al., 2016]{ref28} Zhu, X., Guo, X., Wang, T. and Zhu, L. (2016) \enquote{Dimensionality determination: a thresholding double ridge ratio criterion.} \url{http://cn.arxiv.org/pdf/1608.04457}

		


	\end{description}
\end{singlespace}

\newpage
\begin{table}\tabcolsep=7pt
	\caption{\linespread{1.15}Distribution of $\hat K-K$ with $K=11$ for various detection algorithms under CP model.}
	\small
	\bigskip
	\noindent\makebox[\textwidth][c]{
		\linespread{1.5}\selectfont
		\begin{tabular}{ccccccccccccccccccccc}
			\hline
			& & & & &$\hat K-K$& & & \smallskip \\  \cline{3-9}
			Scenarios & Procedures        &$\le -3$&$-2$&-1&0&1&2&$\ge 3$&  \smallskip \\ \hline
			
			(i)$\sigma=1$
			&  cumSeg            &0&0&2&873&119&6&0&\\
			&   WBS                &0 &0&0&982&16&2&0&\\
			&  SMUCE           &0&0&0&995&5&0&0\\
			&  TMAR                 &0&0&0&998&2&0&0&\medskip\\
			\hline
			
			(ii)$\sigma^2=3$
			&    cumSeg            &736&234&27&3&0&0&0&\\
			&   WBS                &255 &598&135&12&0&0&0&\\
			&  SMUCE           &485&487&28&0&0&0&0&\\
			&  TMAR                 &0&2&46&645&264&39&4&\\
			
			\hline
			
			(iii)
			&    cumSeg            &21&707&236&30&6&0&0&\\
			&   WBS                &0 &422&457&119&2&0&0&\\
			&  SMUCE           &0&342&488&170&0&0&0&\\
			&  TMAR                 &0&0&18&859&121&2&0&\\
			
			\hline
			
			(iv)
			&    cumSeg            &1000&0&0&0&0&0&0&\\
			&   WBS                &0 &0&0&0&1&5&993&\\
			&  SMUCE           &0&0&0&0&0&0&1000&\\
			&  TMAR                 &1&8&66&331&384&164&46&\\
			
			\hline
	\end{tabular}}
	\label{t4}
\end{table}

\newpage
\begin{table}\tabcolsep=7pt
	\caption{\linespread{1.15}Distribution of $\hat K-K$ for various detection algorithms under CP alternative model.}
	\small
	\bigskip
	\noindent\makebox[\textwidth][c]{
		\linespread{1.5}\selectfont
		\begin{tabular}{ccccccccccccccccccccc}
			\hline
			& & & & &$\hat K-K$& & & \smallskip \\  \cline{3-9}
			Scenarios & Procedures        &$\le -3$&$-2$&-1&0&1&2&$\ge 3$&  \smallskip \\ \hline
			
			(i)$\sigma=1$
			&  cumSeg      &237&262&251&150&94&6&0&\\
			&   WBS                &7 &49&215&599&128&2&0&\\
			&  SMUCE           &18&128&535&318&1&0&0&\\
			&  TMAR         &0&0&32&899&68&1&0&\medskip\\
			\hline
			
			(ii)$\sigma^2=3$              &    cumSeg &1000&0&0&0&0&0&0&\\
			&   WBS                &1000&0&0&0&0&0&0&\\
			&  SMUCE           &1000&0&0&0&0&0&0&\\
			&  TMAR                 &2&15&92&262&314&210&105&\\
			
			\hline
			
			(iii)        &    cumSeg            &999&1&0&0&0&0&0&\\
			&   WBS                &997&3&0&0&0&0&0&\\
			&  SMUCE           &1000&0&0&0&0&0&0&\\
			&  TMAR                 &1&8&90&466&320&95&20&\\
			
			\hline
			
			(iv)              &    cumSeg            &997&3&0&0&0&0&0&\\
			&   WBS                &0 &0&0&1&1&6&992&\\
			&  SMUCE           &0&0&0&0&0&0&1000&\\
			&  TMAR                 &0&8&72&553&297&66&4&\\
			
			\hline
	\end{tabular}}
	\label{t3}
\end{table}

\newpage
\begin{table}\tabcolsep=7pt
	\caption{\linespread{1.15}Distribution of $\hat K-K$  using various detection algorithms under CP model for variance change.}
	\small
	\bigskip
	\noindent\makebox[\textwidth][c]{
		\linespread{1.5}\selectfont
		\begin{tabular}{ccccccccccccccccccccc}
			\hline
			& & & & &$\hat K-K$& & & \smallskip \\  \cline{3-9}
			Scenarios & Procedures        &$\le -3$&$-2$&-1&0&1&2&$\ge 3$&  \smallskip \\ \hline
			
			(i)$\sigma=1$
			&  SMUCE          &1000&0&0&0&0&0&0&\\
			&   BS                &1000 &0&0&0&0&0&0&\\
			&  PELT             &17&2&0&981&0&0&0\\
			&  TMAR            &0&&1&919&80&0&0&\medskip\\
			\hline
			
			(ii)$\sigma^2=3$
			&    SMUCE        &1000&0&0&0&0&0&0&\\
			&   BS                &1000&0&0&0&0&0&0&\\
			&  PELT             &6&0&303&691&0&0&0&\\
			&  TMAR            &0&0&96&903&1&0&0&\\
			
			\hline
			
			(iii)              &    SUMCE            &1000&0&0&0&0&0&0&\\
			&   BS                &1000&0&0&0&0&0&0&\\
			&  PELT           &1000&0&0&0&0&0&0&\\
			&  TMAR                 &0&0&0&229&619&142&10&\\
			
			\hline
			
			(iv)              &    SMUCE           &1000&0&0&0&0&0&0&\\
			&   BS                &1000&0&0&0&0&0&0&\\
			& PELT          &2&5&2&79&69&173&670&\\
			&  TMAR                 &0&1&36&613&293&55&2&\\
			
			\hline
	\end{tabular}}
	\label{t2}
\end{table}

\end{document}